\documentclass[letterpaper, 10 pt, conference]{ieeeconf}
\usepackage{comment}

\IEEEoverridecommandlockouts                              

\usepackage{graphics} 
\usepackage{epsfig} 
\usepackage{mathptmx} 
\usepackage{times} 
\usepackage{amsmath} 
\usepackage{amssymb}  
\usepackage{epstopdf}
\usepackage{xcolor}
\usepackage{cite} 
\usepackage[margin=19.1mm]{geometry}
\providecommand{\keyword}[1]
{
  \small	
  \textbf{\textit{Keywords---}} #1
}

\title{\LARGE \bf 
Solving Infinite-Horizon Optimal Control Problems using the Extreme Theory of Functional Connections}

\author{Tanay Raghunandan Srinivasa$^{1}$, Suraj Kumar$^{2}$
\thanks{$^{1}$Tanay Raghunandan Srinivasa is a BTech Student of Robotics \& Autonomous Systems,
        Plaksha University, Mohali, Punjab, India-140306
        {\tt\small tanay.srinivasa@plaksha.edu.in}}%
\thanks{$^{2}$Suraj Kumar is with the U R Rao Satellite Center, Indian Space Research Organisation,
        Bengaluru, Karnataka, India
        {\tt\small surajk@ursc.gov.in}}%
}

\begin{document}

\maketitle
\thispagestyle{empty}
\pagestyle{empty}

\begin{abstract}
This paper presents a physics-informed machine learning approach for synthesizing optimal feedback control policy for infinite-horizon optimal control problems by solving the Hamilton–Jacobi–Bellman (HJB) partial differential equation(PDE). The optimal control policy is derived analytically for affine dynamical systems with separable and strictly convex control costs, expressed as a function of the gradient of the value function. The resulting HJB-PDE is then solved by approximating the value function using the Extreme Theory of Functional Connections (X-TFC)—a hybrid approach that combines the Theory of Functional Connections (TFC) with the Extreme Learning Machine (ELM) algorithm. This approach ensures analytical satisfaction of boundary conditions and significantly reduces training cost compared to traditional Physics-Informed Neural Networks (PINNs). We benchmark the method on linear and non-linear systems with known analytical solutions as well as demonstrate its effectiveness on control tasks such as spacecraft optimal de-tumbling control. \\
\end{abstract}
\keyword{\small Optimal Control, Physics Informed Neural Networks, Extreme Learning Machines}

\section{Introduction}
Optimal control problems (OCPs) aim to determine the inputs to a dynamical system that optimize a given performance index while satisfying constraints on the system dynamics. For many real-time applications, it is essential to derive optimal feedback policies that can generate optimal behavior from arbitrary initial conditions. Due to the complexity of real-world systems, optimal control problems are typically solved using numerical methods, except in simple cases such as the Linear Quadratic Regulator. One of the most widely used approaches is direct trajectory optimization, which transcribes the infinite-dimensional optimal control problem into a finite-dimensional nonlinear program\cite{rao2009survey}. However, these methods yield open-loop solutions, optimal trajectories corresponding to specific initial conditions, rather than general feedback control laws. One approach to obtaining a feedback control law from trajectory optimization is to solve the underlying OCP over a receding horizon and apply only the first control input at each step—an approach known as Model Predictive Control (MPC)\cite{garcia1989model,rawlings2017model}. Convex formulations of MPC have been successfully applied in many control applications, largely because convex programs can be solved efficiently in polynomial time. However, this approach is generally limited to systems with moderate dimensionality. For high-dimensional systems, MPC is often implemented using reduced-order models that approximate the full system dynamics\cite{10883705}. 

Another common approach to feedback control laws is to leverage Bellman’s Principle of Optimality, which leads to the Hamilton–Jacobi–Bellman (HJB) partial differential equation (PDE)—a necessary and sufficient condition for optimality\cite{bertsekas2012dynamic}. By solving the HJB-PDE in principle, one can synthesize optimal feedback control laws that can be directly applied to achieve optimal system behavior. For a long time, solving the HJB equation was limited to low-dimensional systems due to the curse of dimensionality. However, the advent of machine learning techniques for solving PDEs has opened new research avenues under the umbrella of Physics-Informed Machine Learning (PIML), enabling, in principle, the learning of optimal feedback control laws for high-dimensional systems.

The core idea behind Physics-Informed Machine Learning (PIML) is to use neural networks (NNs) as universal function approximators, where the network is trained using the governing PDE as part of the loss function. As a result, the NN inherently satisfies the underlying physics of the problem. The development of Physics-Informed Neural Networks (PINNs) was significantly advanced by the advent of automatic differentiation techniques\cite{osti_1595805}.
For a PINN to satisfy the PDE, it must both satisfy the PDE as well its corresponding boundary condition. Hence traditional PINNs trained to solve PDEs use two different losses which are linearly combined during the training process \cite{SIRIGNANO20181339}. While simpler PDEs can be solved with these multiple objectives, gradient imbalance issues arise when solving larger problems and competing loss function affects accuracy.

The incorporation of the Theory of Functional Connections (TFC)—a method for solving PDEs using a free function and a functional—into PIML helps address the problem of multiple objectives\cite{make2010004}. In TFC-based PIML methods, a neural network serves as the free function, and the constructed solution automatically satisfies the boundary conditions of the PDE by design. However, training deep TFC-based PINNs using gradient-based optimization can be computationally expensive\cite{schiassi2020}. To mitigate this, the Extreme Learning Machine (ELM) algorithm is introduced to accelerate the training process. ELM computes the output layer weights of a single hidden layer fully connected network (SHLFCN) analytically, using randomly initialized and fixed input weights\cite{HUANG2006489}. The combination of ELM and TFC leads to the Extreme-TFC (X-TFC) method, where an SHLFCN is first trained using the ELM algorithm to obtain an approximate solution to the PDE, and then optionally refined using gradient-based methods to improve accuracy\cite{schiassi2021extreme,HUANG2006489}. 

The main contribution of this work is the development of an optimal controller for infinite-horizon problems using the HJB PDE, which arises as the necessary and sufficient condition of optimality from the Principle of Optimality applied to the OCP. The optimal policy is first derived analytically as a function of the gradient of the value function, and then the value function is learned using X-TFC by solving the resulting HJB PDE.

The remainder of the paper is organized as follows. Section \ref{method} provides the necessary background, and details the optimal control formulation and training methodologies. Section \ref{results} presents the simulation results. Section \ref{conc} concludes the paper. 

\section{Methodology}
\label{method}
\subsection{Infinite Horizon Optimal Control Problem}
Consider a dynamical system governed by the differential equation:
\begin{equation}
    \dot{\mathbf{x}}(t) = f(\mathbf{x}(t), \mathbf{u}(t)), \quad \mathbf{x}(0) = \mathbf{x}_0,
\end{equation}
where $\mathbf{x}(t) \in \Omega \subseteq \mathbb{R}^n$ is the state vector, $\mathbf{u}(t) \in \mathcal{U} \subseteq \mathbb{R}^m$ is the control input, and $f : \mathbb{R}^n \times \mathbb{R}^m \rightarrow \mathbb{R}^n$ represents the system dynamics.

A control policy is defined as a mapping $\pi : \mathbb{R}^n \rightarrow \mathbb{R}^m$ that prescribes a control input based on the current state, i.e., $\mathbf{u}(t) = \pi(\mathbf{x}(t))$. Given an initial condition $\mathbf{x}_0$, the performance of a policy $\pi$ is evaluated using the infinite-horizon cost functional.
\begin{equation}
    J(\mathbf{x}_0, \pi) = \int_0^{\infty} \ell(\mathbf{x}(t), \pi(\mathbf{x}(t))) \, dt,
\end{equation}
where $\ell(\mathbf{x}, \mathbf{u})$ is a non-negative, continuous running cost function. The value function under policy $\pi$ is defined as:
\begin{equation}
    V^{\pi}(\mathbf{x}_0) = J(\mathbf{x}_0, \pi).
\end{equation}

The goal is to find an optimal policy $\pi^*$ that minimizes the cost for every initial state.
\begin{equation}
    V^*(\mathbf{x}_0) = \inf_{\pi} V^{\pi}(\mathbf{x}_0) = \inf_{\pi} \int_0^{\infty} \ell(\mathbf{x}(t), \pi(\mathbf{x}(t))) \, dt.
\end{equation}

Under suitable regularity conditions, the optimal value function \( V^*(\mathbf{x}) \) satisfies the \textit{Hamilton-Jacobi-Bellman} (HJB) equation.
\begin{equation}
\label{hjb}
    0 = \min_{\mathbf{u} \in \mathcal{U}} \left\{\ell(\mathbf{x}, \mathbf{u}) + \nabla_{x} V^*(\mathbf{x})^\top f(\mathbf{x}, \mathbf{u}) \right\}, \quad V^*(\mathbf{0}) = \mathbf{0}
\end{equation}
where \( \mathcal{U} \) is the admissible control set. The optimal policy \( \pi^*(\mathbf{x}) \) is then given by:
\begin{equation}
    \pi^*(\mathbf{x}) = \arg\min_{\mathbf{u} \in \mathcal{U}} \left\{ \ell(\mathbf{x}, \mathbf{u}) + \nabla_{x} V^*(\mathbf{x})^\top f(\mathbf{x}, \mathbf{u}) \right\}.
\end{equation}

\subsection{HJB Optimality and Optimal Control} 
We consider control affine dynamics with separable cost function for state and control given as:
\begin{equation}
    \dot{\mathbf{x}} = \mathbf{f}(\mathbf{x},\mathbf{u}) = \mathbf{A}(\mathbf{x}) + \mathbf{B}(\mathbf{x})\mathbf{u}    
\end{equation}
\begin{equation}
    \ell(\mathbf{x},\mathbf{u}) = r(\mathbf{x}) + g(\mathbf{u})
\end{equation}
The resultant HJB is given as:
\begin{equation} \label{eqn:hjb-eqn}
    \mathbf{V_x^*}^\top\mathbf{A}(\mathbf{x}) + r(\mathbf{x}) + \inf_{\mathbf{u}\in\mathcal{U}}\{\mathbf{V_x^*}^\top\mathbf{B}(\mathbf{x})\mathbf{u} + g(\mathbf{u})\} = 0
\end{equation}
subject to
\begin{equation}
    V(\mathbf{0}) = \mathbf{0}
\end{equation}
The optimal control policy \( \pi^*(\mathbf{x}) \) is given as:
\begin{equation}
\label{opt_policy}
        \pi^*(\mathbf{x}) = \arg\inf_{\mathbf{u}\in\mathcal{U}}\{\mathbf{V_x^*}^\top\mathbf{B}(\mathbf{x})\mathbf{u} + g(\mathbf{u})\}
\end{equation}
where $\mathbf{V_x^*} = \nabla_{x} V^*(\mathbf{x})$. Eq(\ref{opt_policy}) can be solved analytically if the cost associated with the control term namely $g(\mathbf{u})$ is strictly convex using the properties of convex conjugate function and its gradient\cite{rockafellar1997convex}. The optimal control can be computed as\cite{lutter2020hjb}:

\begin{equation} \label{eqn:optimal-controller}
\begin{aligned}
    \frac{\partial}{\partial \mathbf{u}} \left( g(\mathbf{u}) + \mathbf{u}^\top \mathbf{B}^\top(\mathbf{x}) \mathbf{V}^*_x \right) = \nabla_\mathbf{u} \, g(\mathbf{u}) + \mathbf{B}^\top(\mathbf{x}) \mathbf{V}^*_x = \mathbf{0} \\
    \Rightarrow \quad \mathbf{u}^* = \nabla_\mathbf{u} \, g^*\left( -\mathbf{B}^\top(\mathbf{x}) \mathbf{V}^*_x \right).
\end{aligned}    
\end{equation}
Here, $g^*(w) = \sup_u \left\{ u^\top w - g(u) \right\}$ is the convex conjugate of g and $\nabla g^* = (\nabla g)^{-1}$ due to the strong convexity of g\cite{rockafellar1997convex}. We consider a quadratic cost function of the form:
\begin{equation}
    g(\mathbf{u}) = \mathbf{u}^\top\mathbf{R}\mathbf{u}
\end{equation}
The form of $g^*$ changes based on whether we consider a constrained set of control inputs, $\mathbf{u}$. For an unconstrained control space, the optimal policy is given as:
\begin{equation}
    \pi^*(\mathbf{x})_{unconstr} = -\frac{1}{2}\mathbf{R}^{-1}\mathbf{B}^\top(\mathbf{x})\mathbf{V}^*_x
\end{equation}
The constrained optimal policy is given as:
\begin{equation}
   \begin{aligned}
    \pi^*(\mathbf{x})_{constr} &= \nabla g^*(-\mathbf{B}^\top(\mathbf{x})\mathbf{V_x})\\
    &= \begin{cases}
        w_i + \gamma_i, & |w_i| < \alpha_i \\
        (\text{sgn}(w_i)*\alpha_i) + \gamma_i, & otherwise
    \end{cases} \\
    \alpha_i &= \frac{u_{max, i} - u_{min,i}}{2} \\
    \gamma_i &= \frac{u_{max,i} + u_{min,i}}{2}
\end{aligned} 
\end{equation}

where, $\alpha_i$, and $\gamma_i$ are constants determined by the range of the constrained controller $u_i$, and $w = -\mathbf{B}^\top(\mathbf{x})\mathbf{V_x}$.

\subsection{Learning the Value Function using the Extreme Theory of Functional Connections}
The Extreme Theory of Functional Connections (X-TFC) is a hybrid framework for solving parametric differential equations that combines the classical Theory of Functional Connections (TFC) with the representational capabilities of neural networks. Consider a general parametric partial differential equation (PDE) given as:
\begin{equation}
\alpha(t, x)\, \frac{\partial y}{\partial t} + \mathcal{N}[y; \theta(t, x)] = S(t, x), \quad x \in \mathcal{D} \subset \mathbb{R}^n
\end{equation}
subject to initial and boundary conditions. Here, \( \alpha(t, x) \in \mathbb{R} \) is a known scalar parameter field, \( \theta(t, x) \in \mathbb{R}^m \) denotes a set of parameter fields (e.g., coefficients of differential operators), \( \mathcal{N}[y; \theta] \) is a linear or nonlinear differential operator acting on \( y \), and \( S(t, x) \) is a known forcing function.

According to the TFC, the unknown (or latent) solution of the differential equation is approximated using a constrained expression which consists of two parts: a function that analytically satisfies the given constraints, and a functional term that incorporates a freely chosen function. The constrained expression is given as:
\begin{equation}
y(t, x; \Theta) = C(t, x) + D(t, \eta(t,x);\Theta)
\end{equation}
Here, \( C(t, x) \) is analytically constructed function that satisfies all constraints on initial and boundary conditions; \( D(t, x) \) is the null space-like function which projects the free function $\eta(t,x)$ onto space of functions that vanish at constraints and \( \Theta \): learnable parameters of the free function. This construction guarantees that any choice of \( \eta \) leads to a solution \( y(t, x; \Theta) \) that satisfies the initial and boundary conditions analytically. The constrained expression for solving the steady state HJB-PDE is given as: 
\begin{equation}
    V^*(\mathbf{x}) \approx V(x;\Theta) = \eta(\mathbf{x};\Theta) + (V(\mathbf{0}) - \eta(\mathbf{0};\Theta))
\end{equation}
where, $\eta(\mathbf{x};\Theta)$ is the free function approximated using a neural network and $\Theta$ is the parameter of the neural network. With this choice of constrained expression, boundary conditions are analytically satisfied irrespective of the choice of neural network.

In the classical TFC formulation, the free function is typically represented as a linear combination of orthogonal basis functions, such as Legendre or Chebyshev polynomials. In X-TFC, the free function is represented by a neural network, specifically a single layer fully connected feed forward NN, in particular, an Extreme Learning Machine\cite{schiassi2020extremetheoryfunctionalconnections} given as:
\begin{equation} \label{eqn:elm-summation}
\eta(t,x) = \sum_{j=1}^{N} \beta_j\, \sigma\left(w_j^\top x + b_j\right)
\end{equation}
where, N is the number of nodes in the hidden layer, $w_j$ are the input layer weights, $b_j$ are the input layer biases and $\sigma(\cdot)$ is the activation function. According to the ELM algorithm \cite{schiassi2020extremetheoryfunctionalconnections}, the input weights and biases are randomly initialized and remain fixed during training, making them known parameters and only the output layer weights are trained analytically by a solving set of linear equations. Eqn. \eqref{eqn:elm-summation} can be re-written as a tensor operation, given as:
\begin{equation}
    \mathbf{H}\beta = \mathbf{T}
\end{equation}
where, 
\begin{equation}
    \mathbf{H} = \begin{bmatrix}
        \sigma(w_1x_1+b_1) & \cdots & \sigma(w_Nx_1+b_N)\\
        \vdots & \ddots & \vdots \\
        \sigma(w_1x_N+b_1) & \cdots & \sigma(w_Nx_N+b_N)
    \end{bmatrix}
\end{equation}
and, $\mathbf{T}$ is the target output for the network. With this form, the values for $\beta$ can be found analytically using the Moor-Penrose generalized inverse of the matrix $\mathbf{H}$:
\begin{equation}
    \beta = \mathbf{H}^\dagger\mathbf{T}
\end{equation}
In the case of X-TFC using ELM algorithm, the network is first trained to satisfy an initial prediction of the value function analytically, after which its trained using the residual of the HJB via gradient descent. In the case of infinite horizon OCPs with LQR cost, it is found that a good approximation to the value function is given as:
\begin{equation}
    V_{init}(\mathbf{x}) = \mathbf{x}^\top\mathbf{Q}\mathbf{x}
\end{equation}
In addition to the value function approximation, regularization is added to the $\mathbf{H}$ matrix during analytical training of the X-TFC. The next part of the training cycle uses gradient based methods to learn the value function such that it satisfies the HJB PDE. The loss used during this stage of training takes the following form:
\begin{equation}
    r^{(i)} =  \mathbf{V}_{\mathbf{x}^{(i)}}^{*\top}\mathbf{T}(\mathbf{x}^{(i)}) + r(\mathbf{x}^{(i)}) - g^*(-\mathbf{B}^\top(\mathbf{x}^{(i)}) \mathbf{V}_{\mathbf{x}^{(i)}})
\end{equation}
\begin{equation}
    \mathcal{L}_{X-TFC} = \frac{1}{N}\sum_{i=1}^{N} (r^{(i)})^2 + \lambda\cdot||\beta||^2_F
\end{equation}
where, $\mathcal{L}_{X-TFC}$ is the loss used to train the network, $r^{(i)}$ is the HJB PDE residual for a training point $\mathbf{x}^{(i)}$ derived using Eqs. \eqref{eqn:hjb-eqn} and \eqref{eqn:optimal-controller}, $N$ is the number of training points, $\lambda$ is a small regularization weight, $\beta$ are the output weights of the X-TFC network, and $||\cdot||_F$ refers to the Frobenius norm. The choice of activation function in the X-TFC model plays an important role during the training process. Dung et al. \cite{dung_6_2023_1803} shows that the Swish family of activation functions provides the most accurate results when used for traditional PINNs. Hence, the same was chosen in our study while training the X-TFC network.

During the gradient-based methods training stage, two different optimizers are used. Adam with an adaptive learning rate scheduler is used, after which L-BFGS is used for the final training step. L-BFGS chosen as it improves the conditioning of the problem without access to the hessian of a problem. Additional hyperparameters during the process of learning the value function using X-TFC are the number of neurons in the hidden layer, number of datapoints used during the training process, the distribution of the training datapoints, the learning rate for weight updates, the number of iterations during the gradient based training phase. Gradient clipping is applied while training to help with the stability of loss reduction. The value at which the gradient is clipped is also hyperparameter. All of these hyperparameters are specific to the OCP being solved, and hence vary based on the problem. The overall training procedure is given in Table \ref{algo_label}.

\begin{table}[hb]
    \centering
    \begin{tabular}{@{}p{1.75cm}p{5cm}@{}}
        \hline
        \textbf{Algorithm 1:} & \textbf{Training the X-TFC Model} \\
        \hline
        
        \textbf{Input:} & Domain  $\Omega$, HJB PDE, and the optimization algorithm to minimize the loss.\\
        
        \textbf{Goal:} & To find optimal X-TFC network parameters $\theta^*$, such that $\eta(\mathbf{x},\theta)$ approximates the solution $V$ of the HJB PDE.\\         
        \textbf{Step 1:} & Initialize the X-TFC network using the the ELM training algorithm to train the X-TFC using the approximate value function. \\
        
        \textbf{Step 2:} & Train the X-TFC network to find $\theta^*$ by Gradient-Based methods using the HJB PDE residual and the optimization algorithm given as input. \\
        \hline
    \end{tabular}
    \caption{Training Algorithm for X-TFC OCP Controller}
    \label{algo_label}
\end{table}

\section{Numerical Results}
\label{results}
To validate the effectiveness of the proposed method, we first present results on benchmark linear and non-linear optimal control problems with known analytical solutions, followed by application to the inverted pendulum and spacecraft de-tumbling control. All results were generated using PyTorch 2.7.0, where the X-TFC models are trained using Metal on a 14 core Apple M1 Pro with 16 GB of RAM.

\subsection{Problem 1: Double Integrator System}

Consider the following OCP:
\begin{equation}
    \min_{u\in\mathcal{U}} J(x) = \frac{1}{2}\int_0^\infty(\mathbf{x}^\top\mathbf{x}+u^2)dt
\end{equation}

subject to
\begin{equation}
\begin{aligned}
    \dot{x}_1 &= x_2 \\
    \dot{x}_2 &= u
\end{aligned}
\end{equation}
where, $x_1 \in [-1, 1]$ and $x_2 \in [-1,1]$ are the states of the system, and $u \in \mathbb{R}$ is the control input.

The exact solution to the problem is given as:
\begin{equation}
\begin{aligned}
    V^*_{exact}(\mathbf{x}) &= \frac{\sqrt{3}}{2}x_1^2+\frac{\sqrt{3}}{2}x_2^2+x_1x_2 \\
    \pi^*_{exact}(x) &= -\sqrt{3}x_2 - x_1 
\end{aligned}
\end{equation}

\begin{figure}
    \centering
    \includegraphics[width=0.35\textwidth]{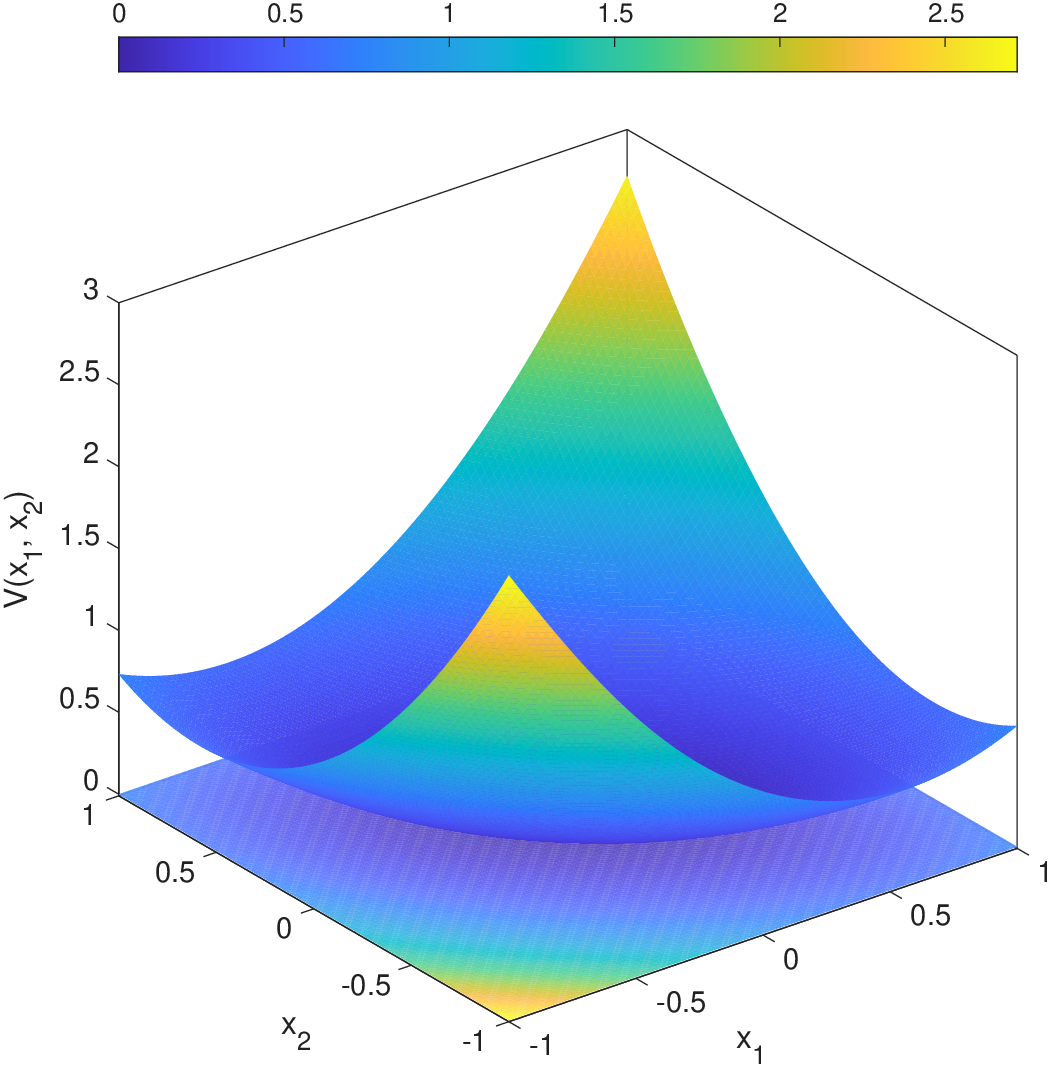}
    \caption{Optimal Value Function of Double Integrator via X-TFC }
    \label{fig:p1_value_fn}
\end{figure}

\begin{figure}
    \centering
    \includegraphics[width=0.35\textwidth]{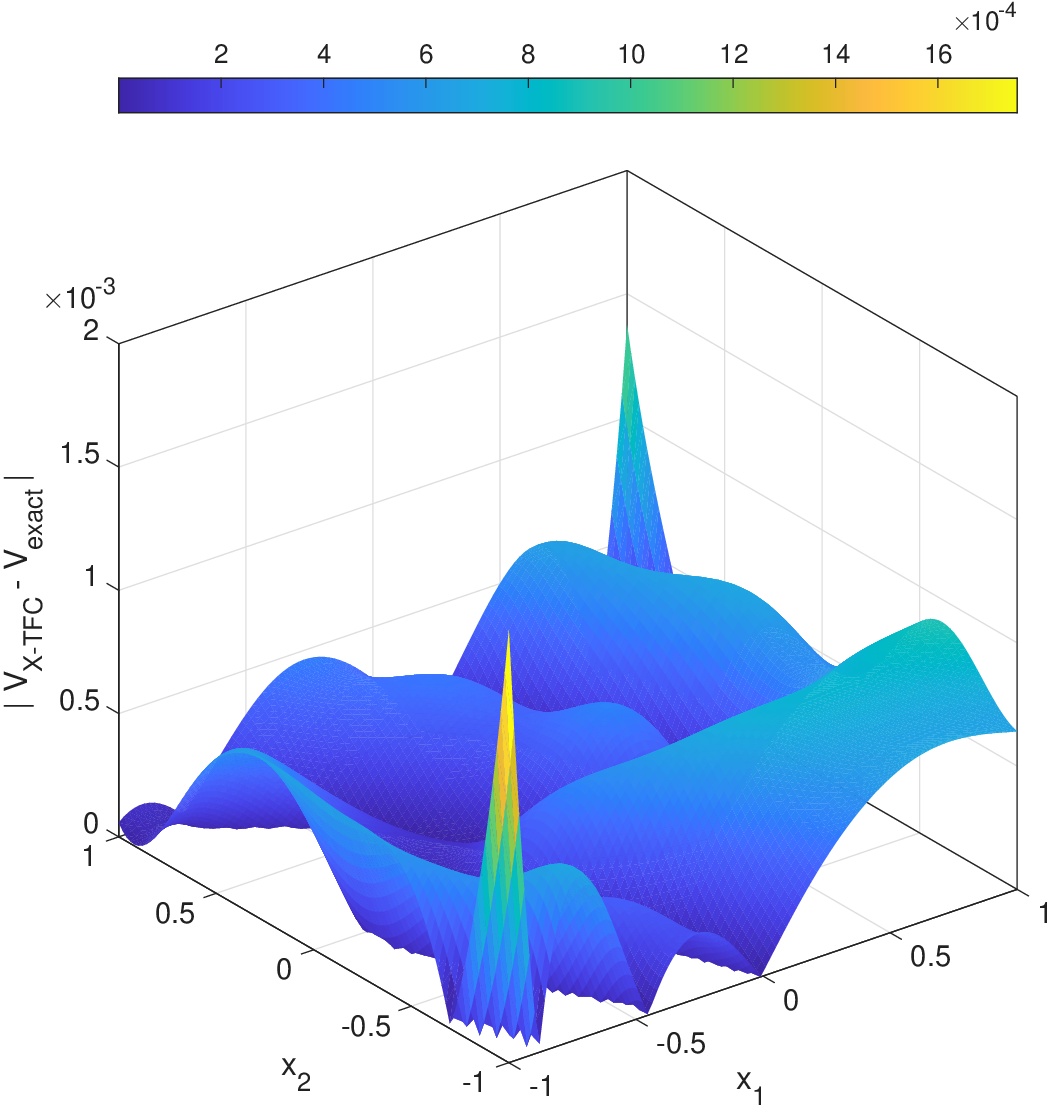}
    \caption{Absolute Error in X-TFC Optimal Value Function and Analytical Solution for Double Integrator}
    \label{fig:p1_residual}
\end{figure}
\begin{figure}
    \centering
    \includegraphics[width=0.40\textwidth]{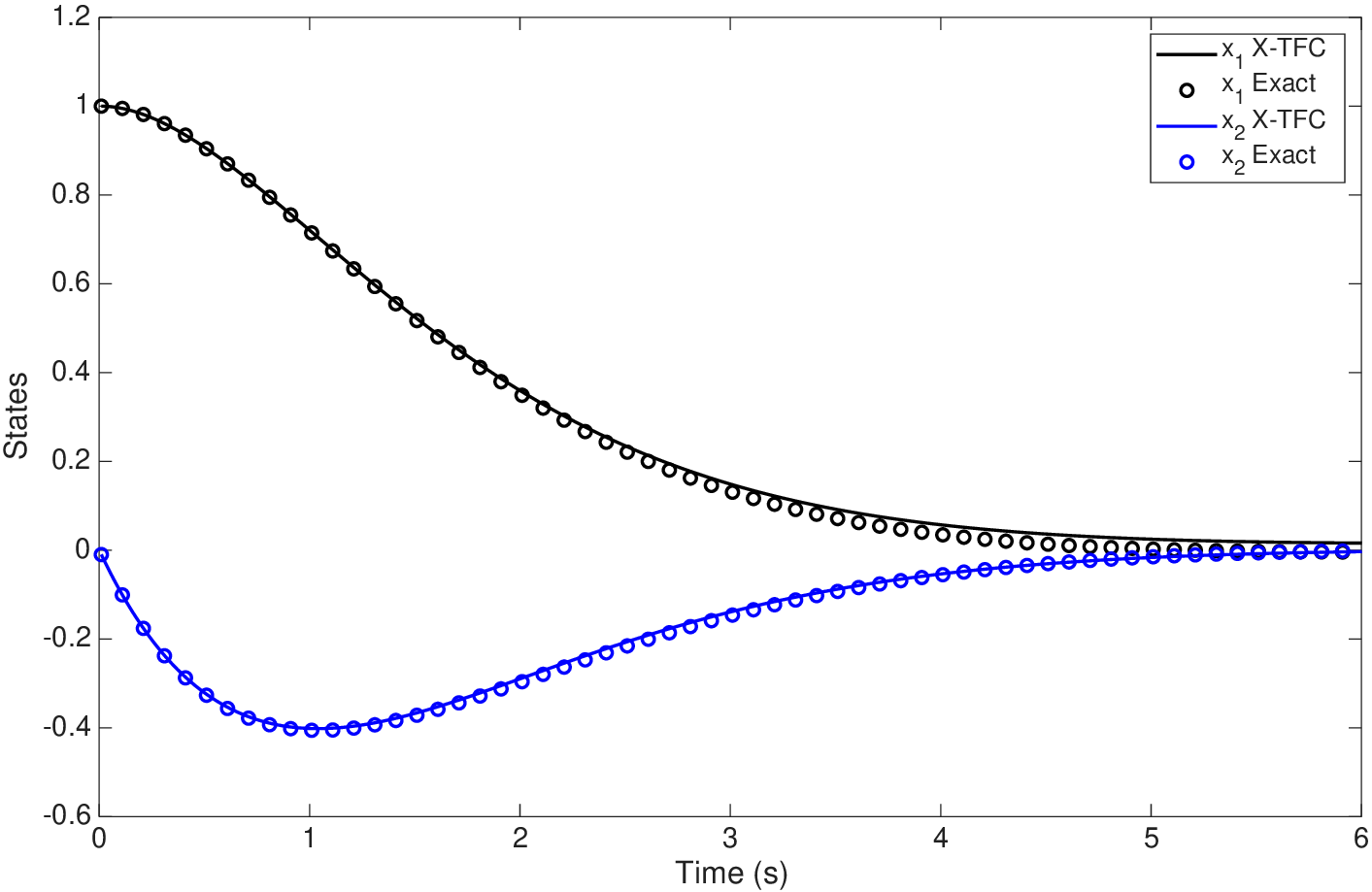}
    \caption{Double Integrator State Trajectory Evolution from arbitrary initial condition under optimal policy}
    \label{fig:p1_state_trajectory}
\end{figure}
\begin{figure}
    \centering    \includegraphics[width=0.40\textwidth]{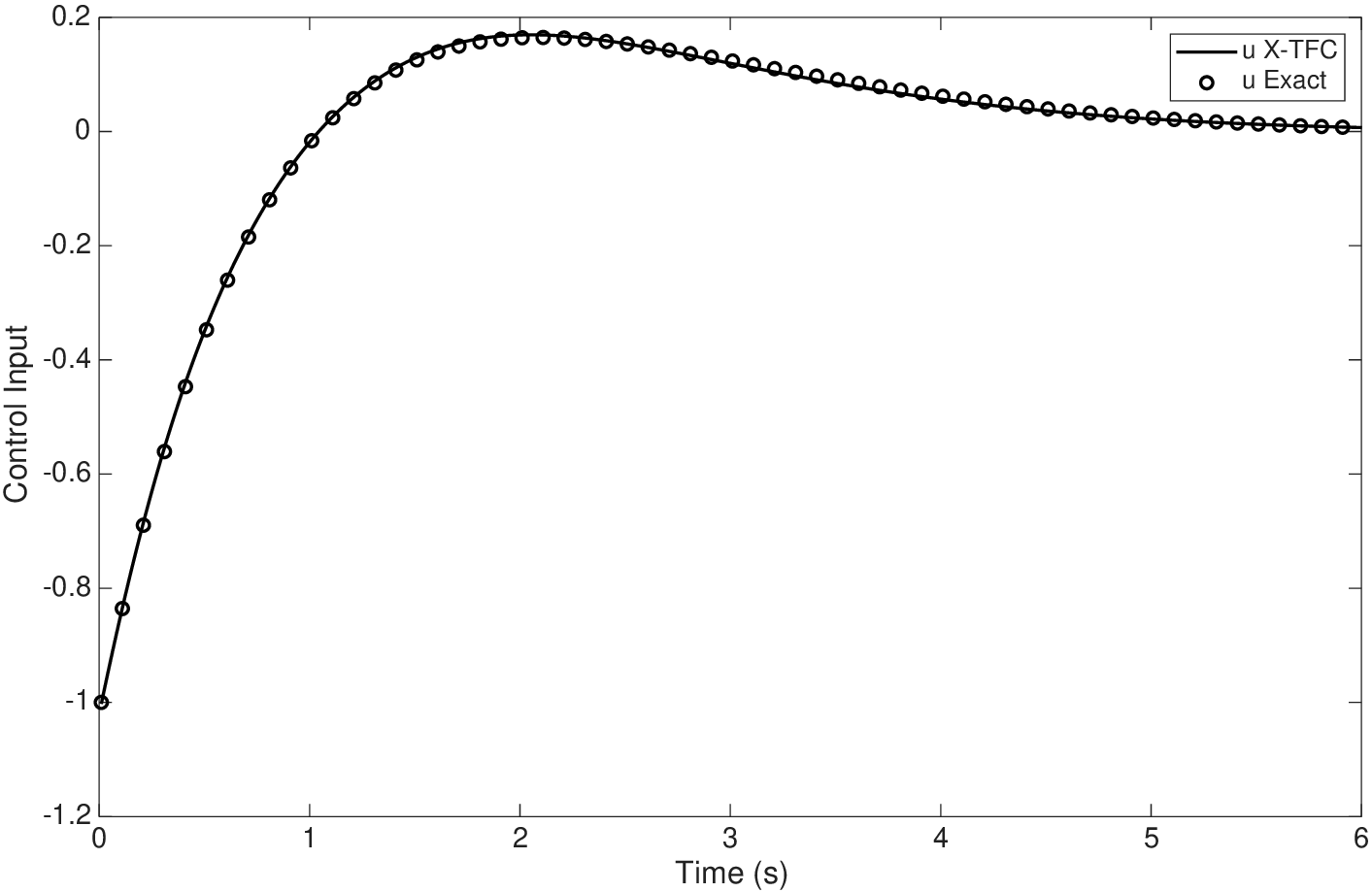}
    \caption{Double Integrator Optimal Control Trajectory}
    \label{fig:p1_control_trajectory}
\end{figure}

The results of the optimal value function for the OCP are given in Fig \ref{fig:p1_value_fn}-\ref{fig:p1_residual}. The observed maximum absolute error in the X-TFC based value function is less than $10^{-3}$. It is noted that the error can be further reduced by training for more epochs. Fig \ref{fig:p1_state_trajectory}-\ref{fig:p1_control_trajectory} shows one sample simulation of state trajectory evolution from arbitrary initial condition under optimal policy. We compare the results from analytical solution as well as X-TFC solution. It is evident that X-TFC closely approximates the analytical solution.   

\subsection{Problem 2: Benchmark Non-Linear Dynamics}

Consider the following OCP:
\begin{equation}
    \min_{u\in\mathcal{U}} J(x,u) = \int_0^\infty(\mathbf{x}^\top\mathbf{x}+u^2)dt
\end{equation}

subject to
\begin{align}
    \dot{x}_1 &= -x_1 + x_2\\
    \dot{x}_2 &= -\frac{1}{2}(x_1+x_2-x_1^2x_2) + x_1u
\end{align}

where, $x_1 \in [-1, 1]$ and $x_2 \in [-1,1]$ are the states of the system, and $u$ is the control input. 

The exact solution to the value function is given as:
\begin{equation}
    V_{exact}^*(\mathbf{x}) = \frac{1}{2}x_1^2 + x_2^2
\end{equation}

\begin{figure}
    \centering
    \includegraphics[width=0.35\textwidth]{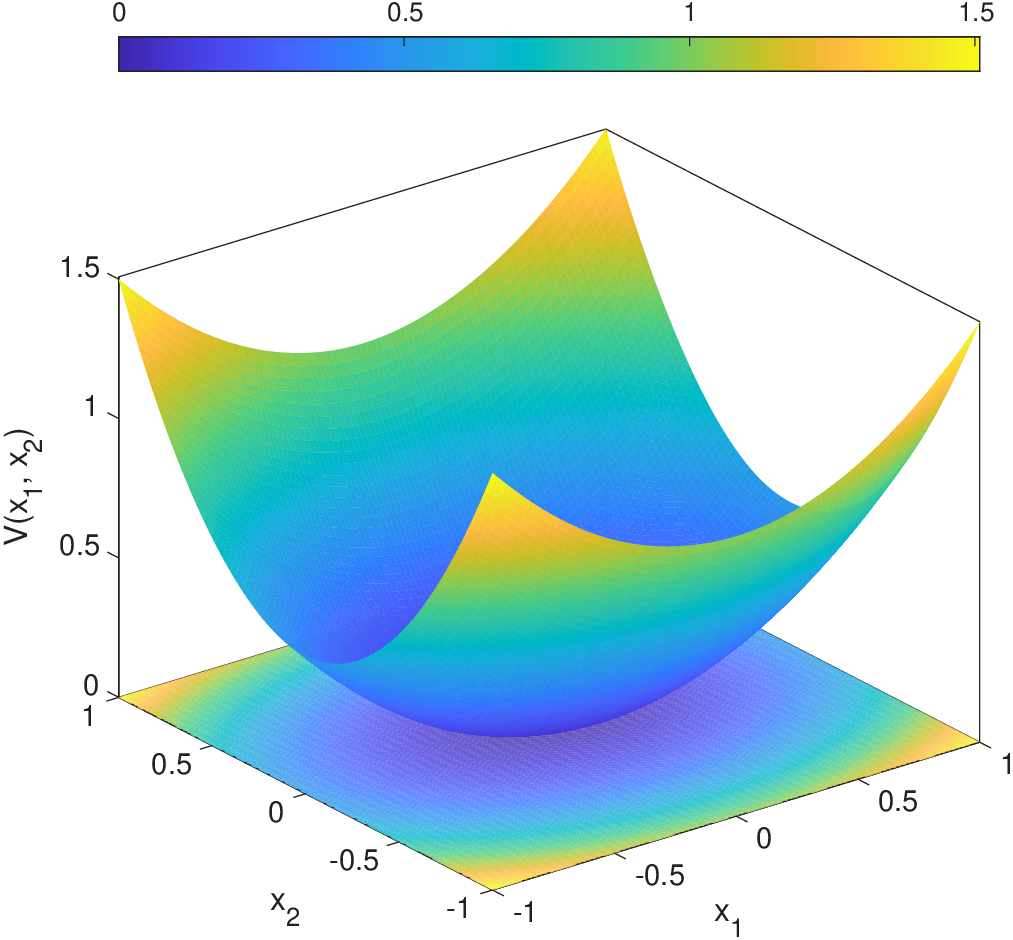}
    \caption{Optimal Value Function for Non-Linear Dynamics via X-TFC}
    \label{fig:p2_value_fn}
\end{figure}

\begin{figure}
    \centering
    \includegraphics[width=0.35\textwidth]{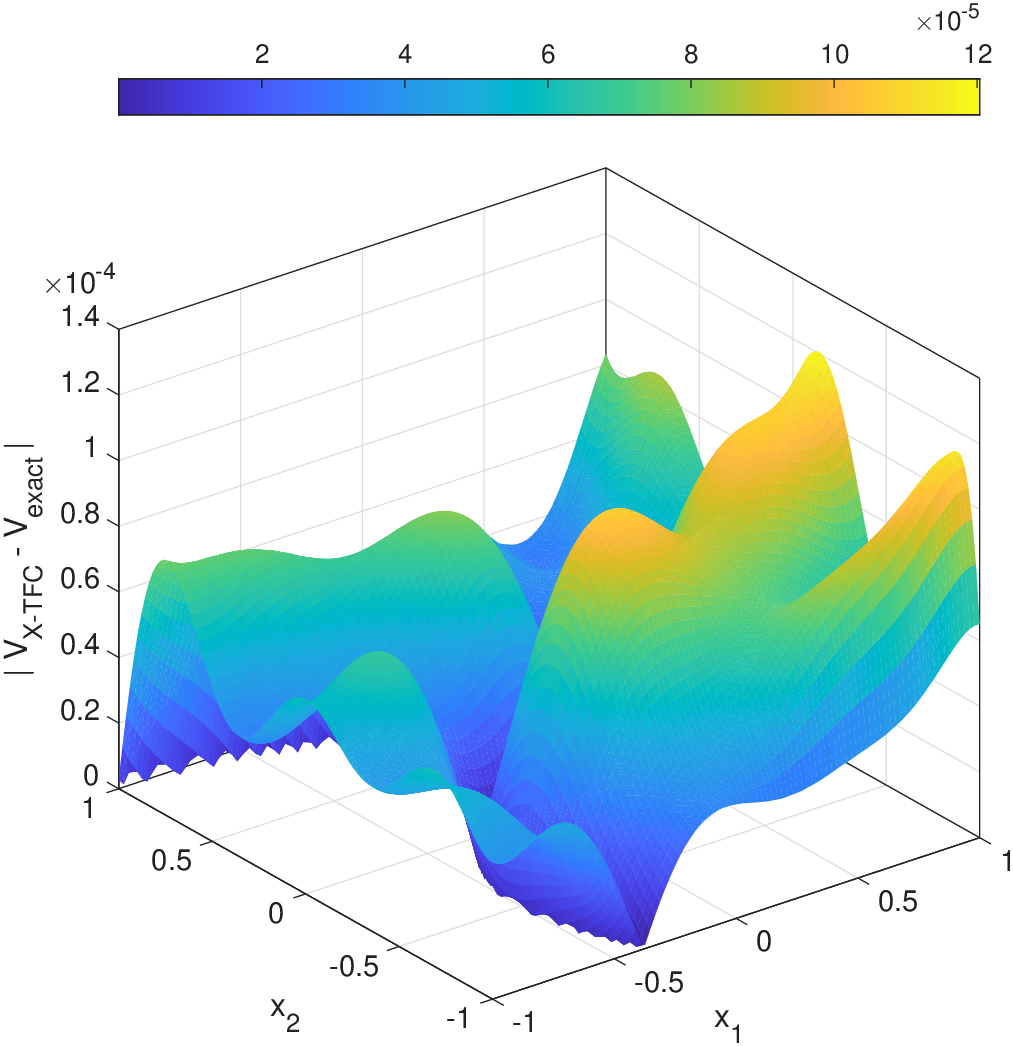}
    \caption{Absolute Error in X-TFC Optimal Value Function and Analytical Solution for Non-Linear Dynamics}
    \label{fig:p2_residual}
\end{figure}

\begin{figure}
    \centering
    \includegraphics[width=0.35\textwidth]{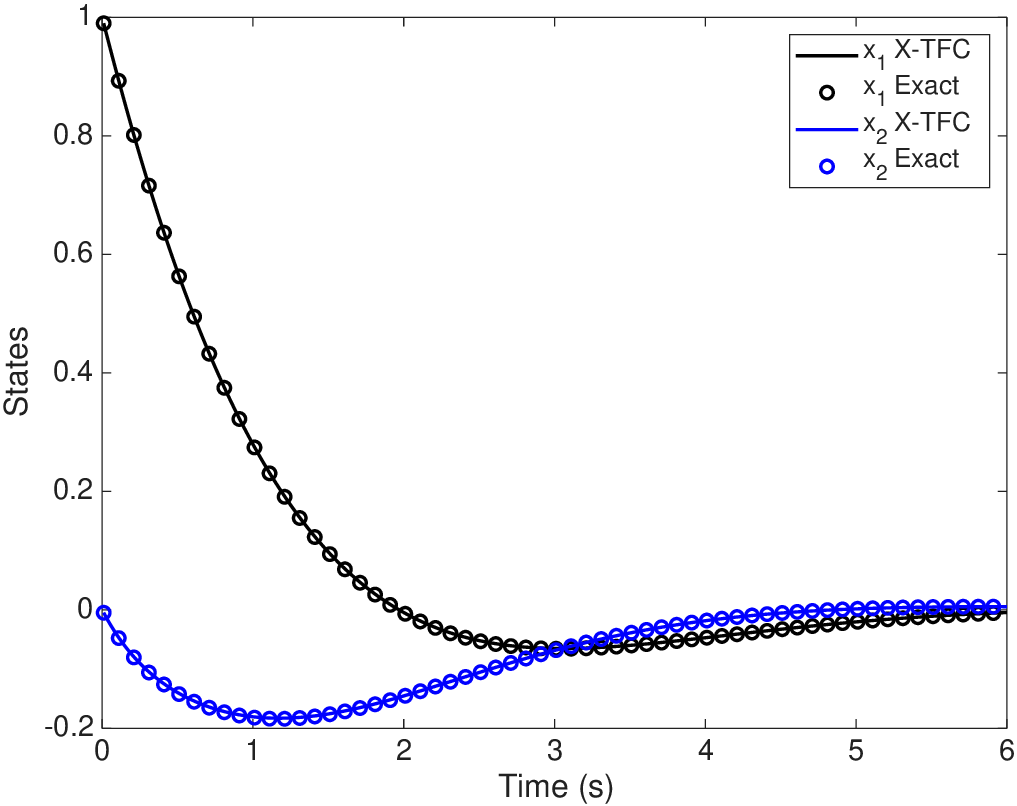}
    \caption{State Trajectory Evolution from arbitrary initial condition under optimal policy for Non-Linear Dynamics}
    \label{fig:p2_state_trajectory}
\end{figure}

\begin{figure}
    \centering
    \includegraphics[width=0.35\textwidth]{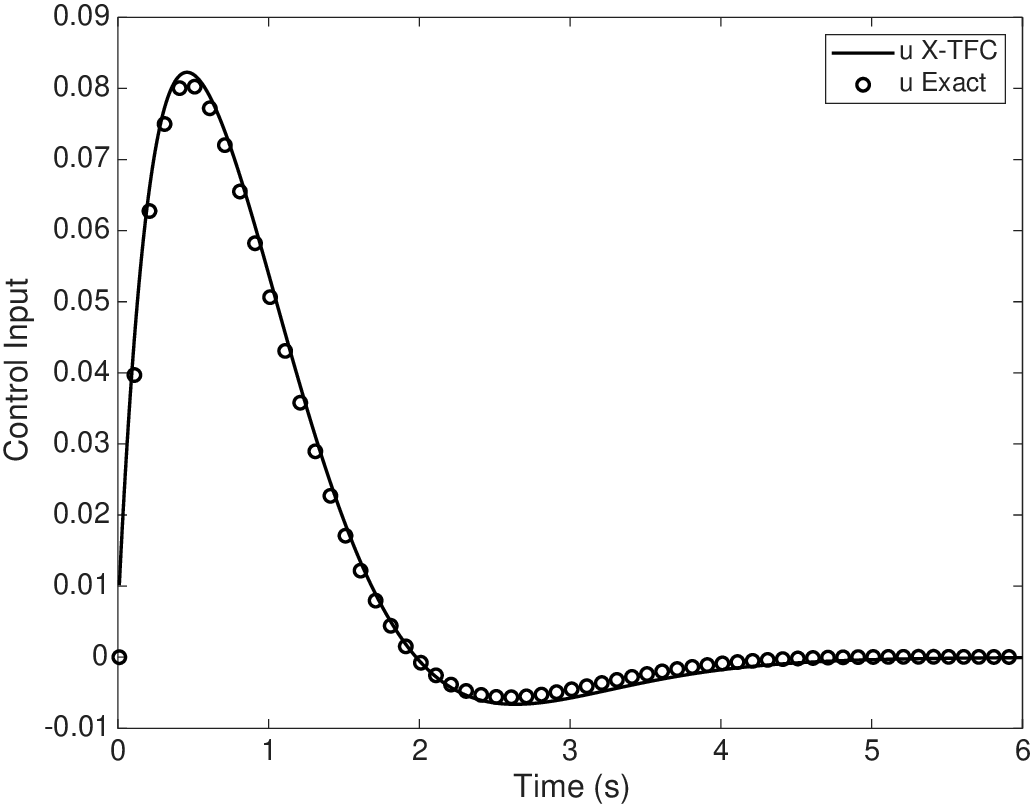}
    \caption{Non-Linear Dynamics Optimal Control Trajectory}
    \label{fig:p2_control_trajectory}
\end{figure}

The results of the optimal value function for the OCP are given in Fig \ref{fig:p2_value_fn}-\ref{fig:p2_residual}. The observed maximum absolute error in the X-TFC based value function is less than $10^{-4}$. Fig \ref{fig:p2_state_trajectory}-\ref{fig:p2_control_trajectory} shows one sample simulation of state trajectory evolution from arbitrary initial condition under optimal policy. We compare the results from analytical solution as well as X-TFC solution. It is evident that X-TFC closely approximates the analytical solution.

\begin{figure}
    \centering
    \includegraphics[width=0.35\textwidth]{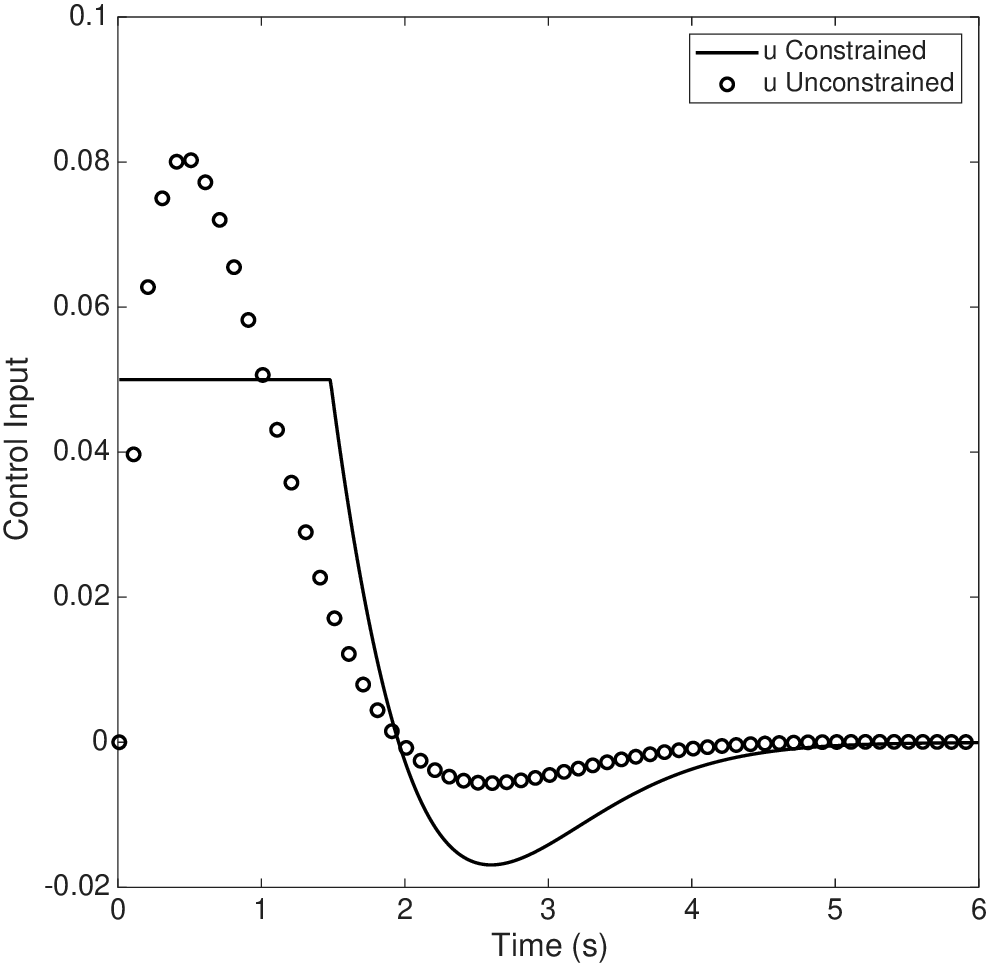}
    \caption{Non-Linear Dynamics Constrained Optimal Control Trajectory}
    \label{fig:p2_control_trajectory_constrained}
\end{figure}

Fig \ref{fig:p2_control_trajectory_constrained} shows one sample simulation of state trajectory evolution from arbitrary initial condition under optimal constrained policy. 

\subsection{Problem 3: Inverted Pendulum}

\begin{figure}
    \centering
    \includegraphics[width=0.40\textwidth]{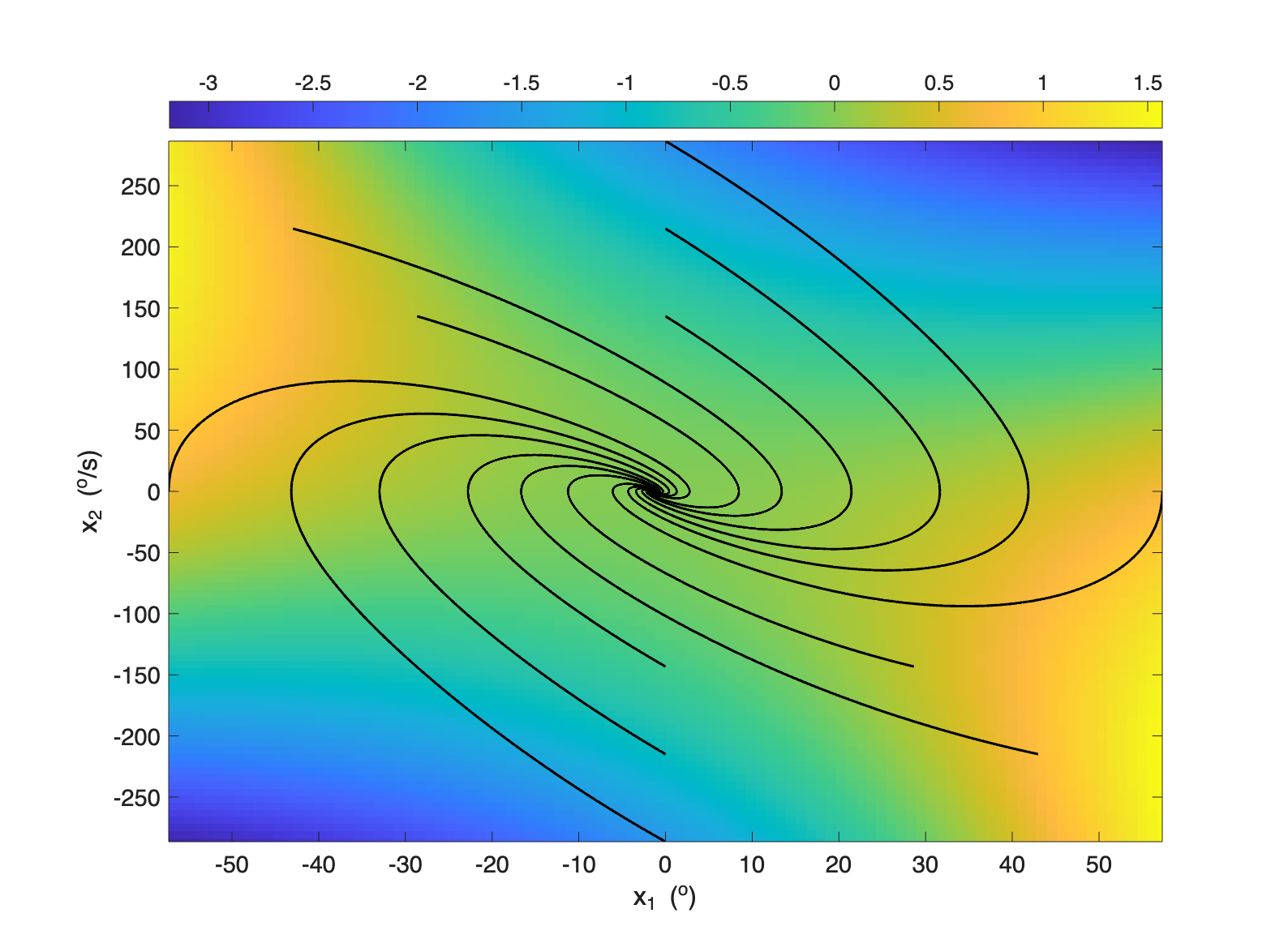}
    \caption{Optimal Value Function Contour and Phase Plot for arbitrary initial condition under optimal policy for the Inverted Pendulum}
    \label{fig:p3_value_fn}
\end{figure}

Consider the following cost function:
\begin{equation}
    \min_{u\in\mathcal{U}} J(x,u) = \int_0^\infty(\mathbf{x}^\top\mathbf{Q}\mathbf{x}+Ru^2)dt
\end{equation}

subject to
\begin{align}
    \dot{x}_1 &= x_2\\
    \dot{x}_2 &= a_2\sin(x_1)+b_2u\\
    a_2 &= \frac{mdg}{I+md^2}\\
    b_2 &= \frac{1}{I+md^2}
\end{align}

where, $x_1$ refers to the angle the pendulum makes with the positive y-axis, $x_2$ refers to the angular rate about the positive k-axis, $u$ is the input torque applied about the positive k-axis, $m$ is the mass of the pendulum, $d$ is the distance of the center of mass (CoM) of the pendulum from the hinge, $I$ is the inertia of the pendulum about its CoM, and $g$ is the acceleration due to gravity. Fig \ref{fig:p3_value_fn} shows the contour of the optimal value function. To evaluate the convergence properties of the optimal controller, Monte Carlo simulations are conducted from a range of initial conditions for the torque-limited pendulum. Fig \ref{fig:p3_value_fn} overlays sample trajectories in the phase plane, each of which converges to the equilibrium point. This demonstrates the global convergence of the learned control policy across the state space.

\subsection{Problem 4: Spacecraft Detumbling}

\begin{figure*}[ht]
  \centering
  \begin{minipage}{.32\textwidth}
    \centering
    \includegraphics[width=\textwidth]{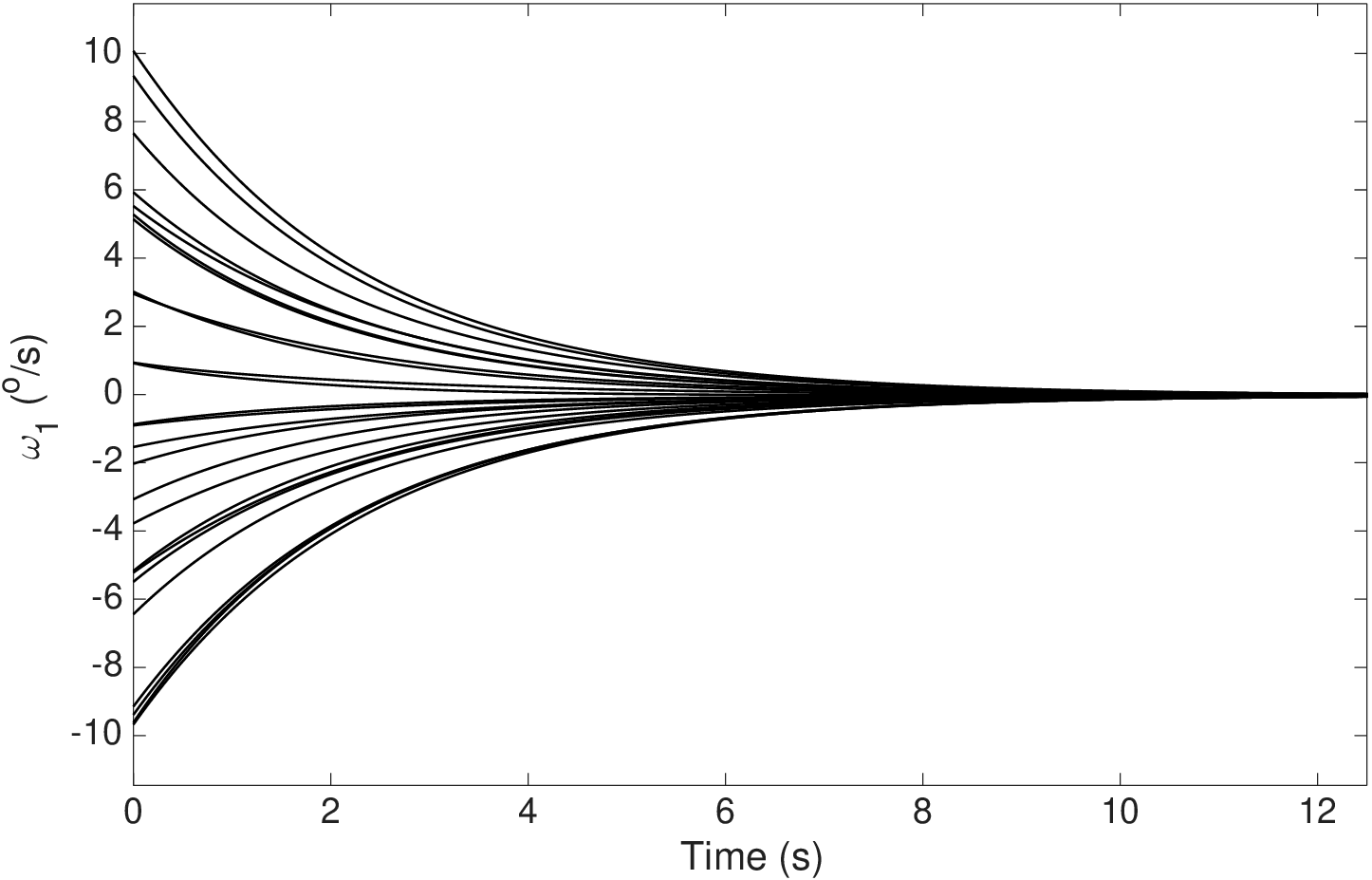}
  \end{minipage}
  \begin{minipage}{.32\textwidth}
    \centering
    \includegraphics[width=\textwidth]{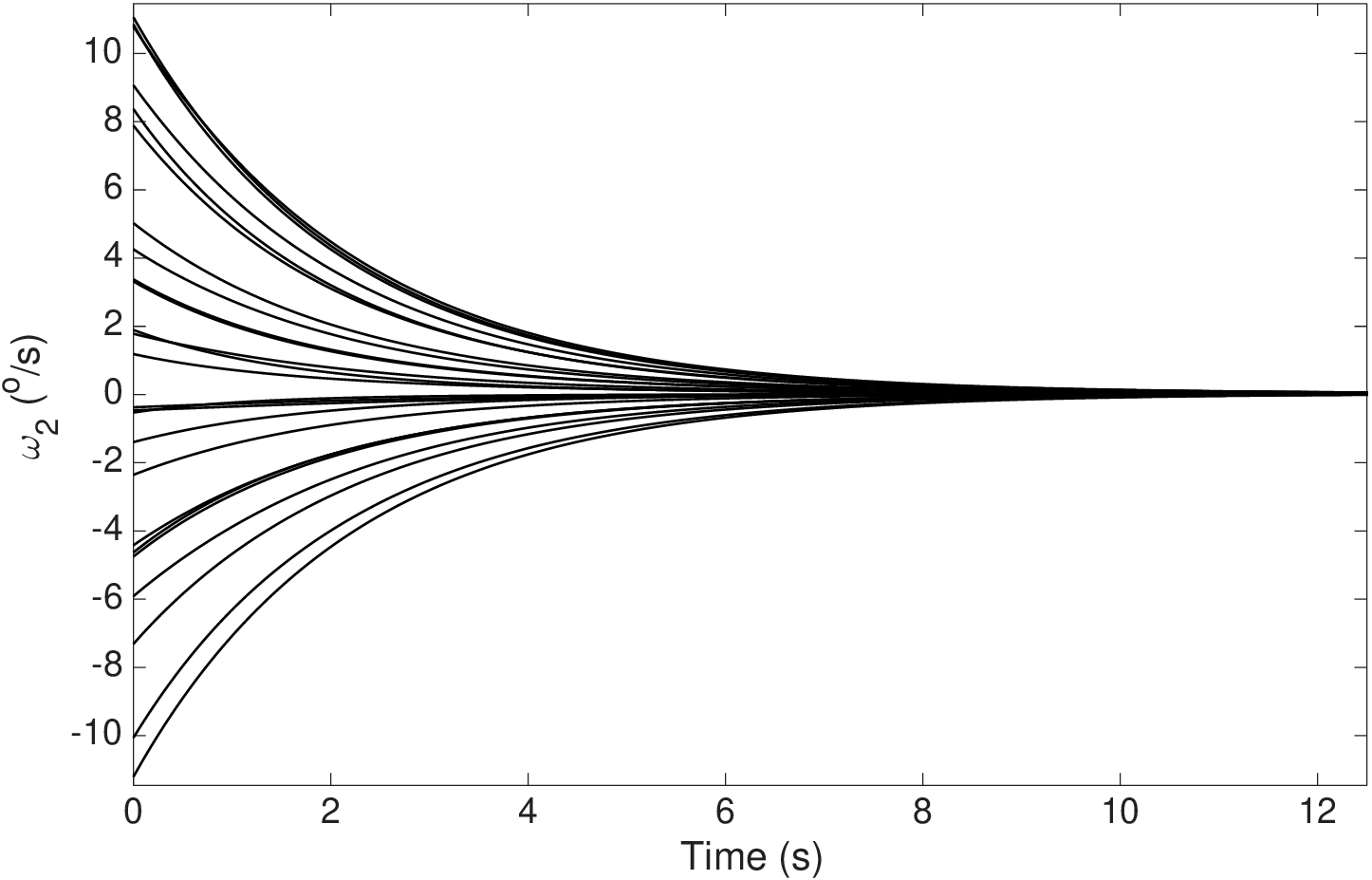}
  \end{minipage}
   \begin{minipage}{.32\textwidth}
    \centering
    \includegraphics[width=\textwidth]{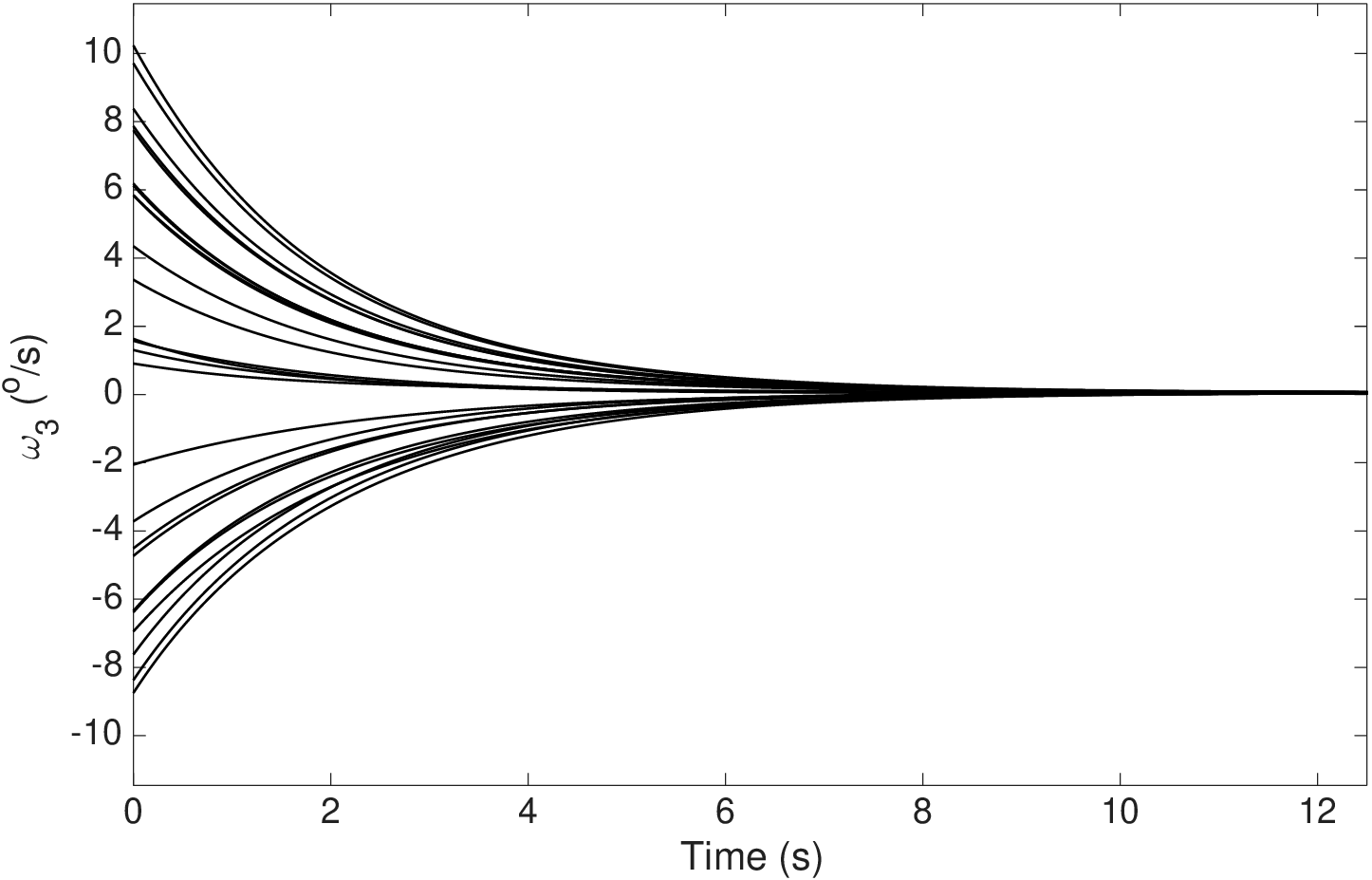}
  \end{minipage}
   \begin{minipage}{.32\textwidth}
    \centering
    \includegraphics[width=\textwidth]{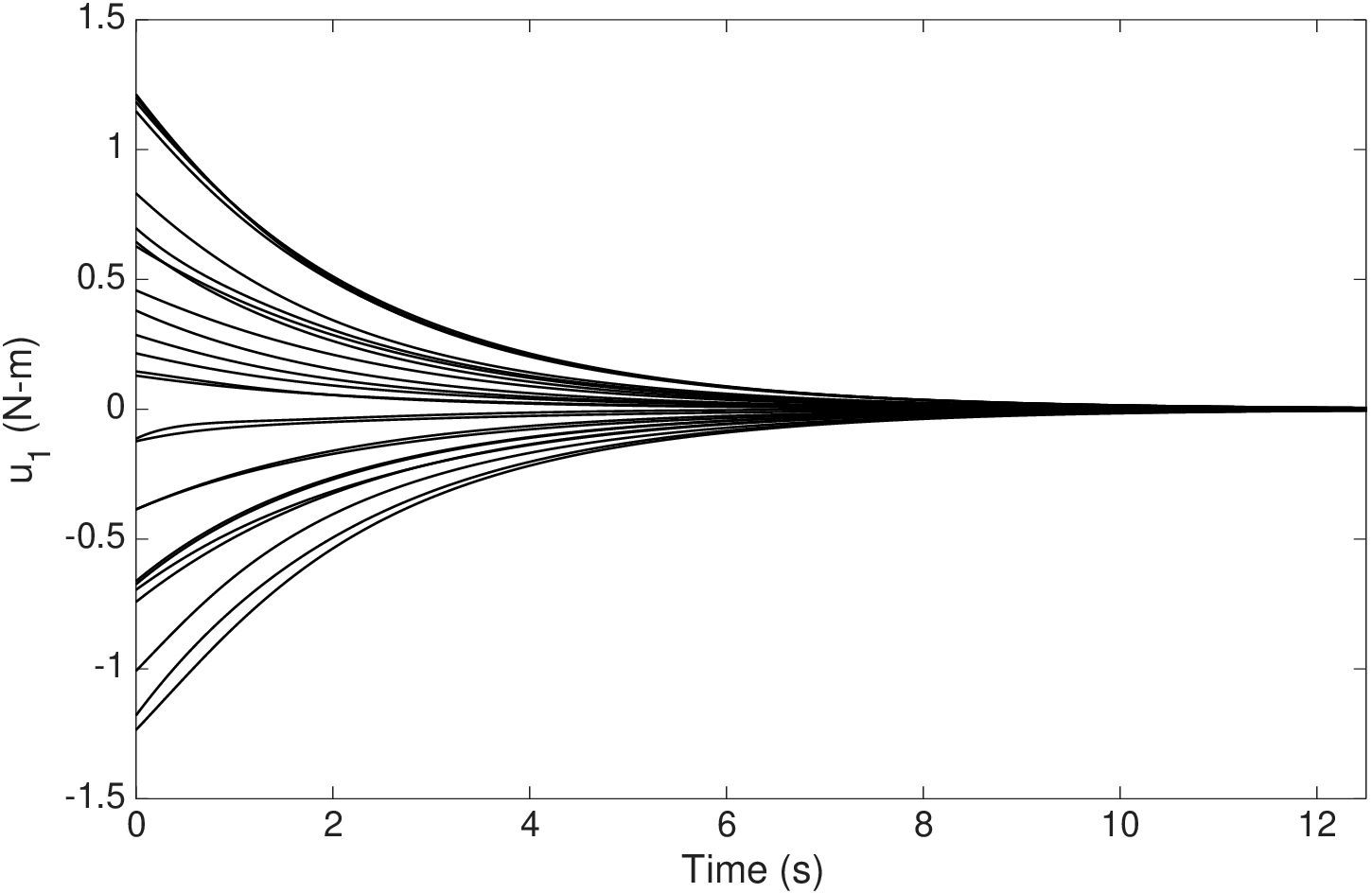}
  \end{minipage}
  \begin{minipage}{.32\textwidth}
    \centering
    \includegraphics[width=\textwidth]{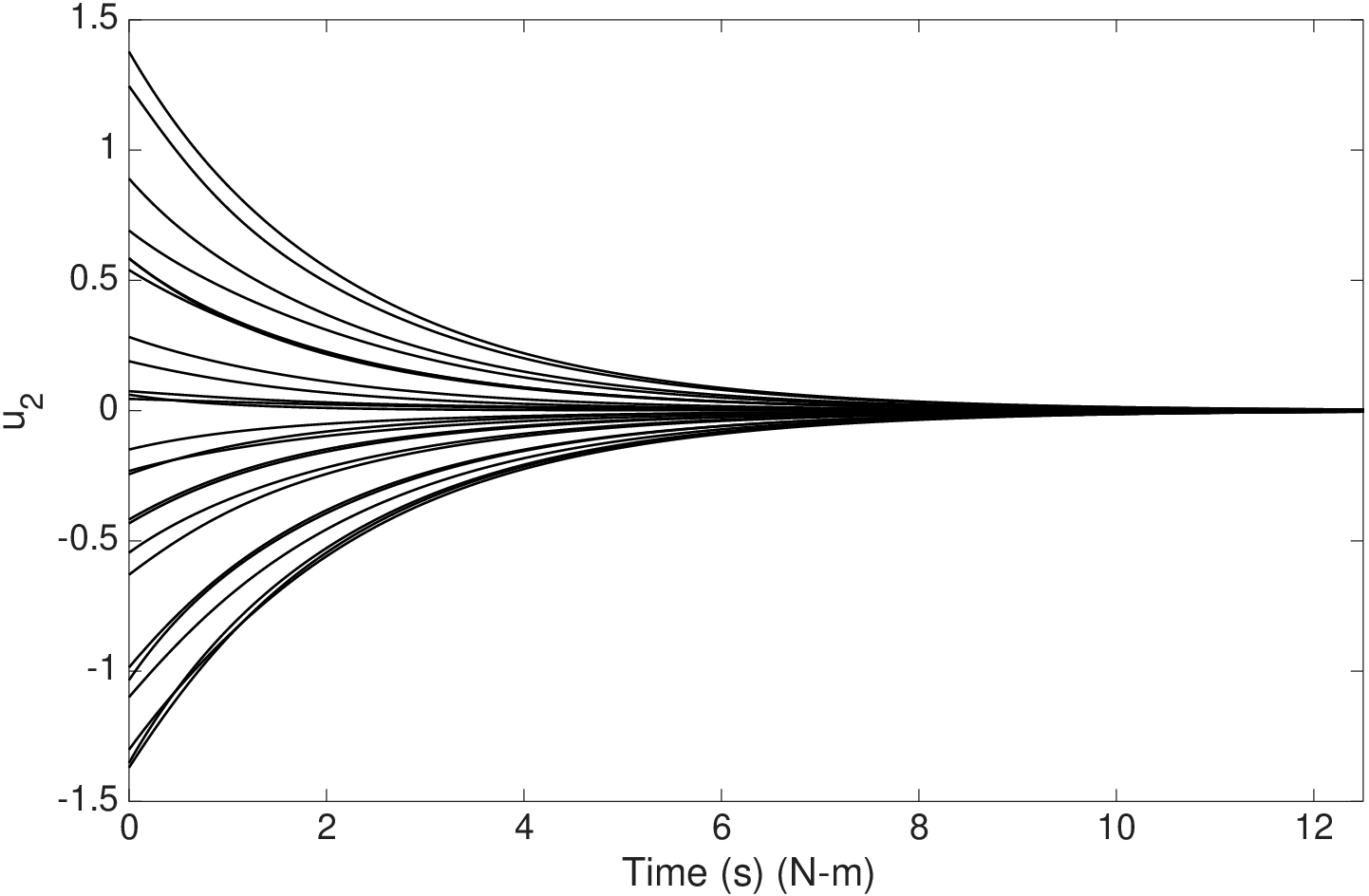}
  \end{minipage}
   \begin{minipage}{.32\textwidth}
    \centering
    \includegraphics[width=\textwidth]{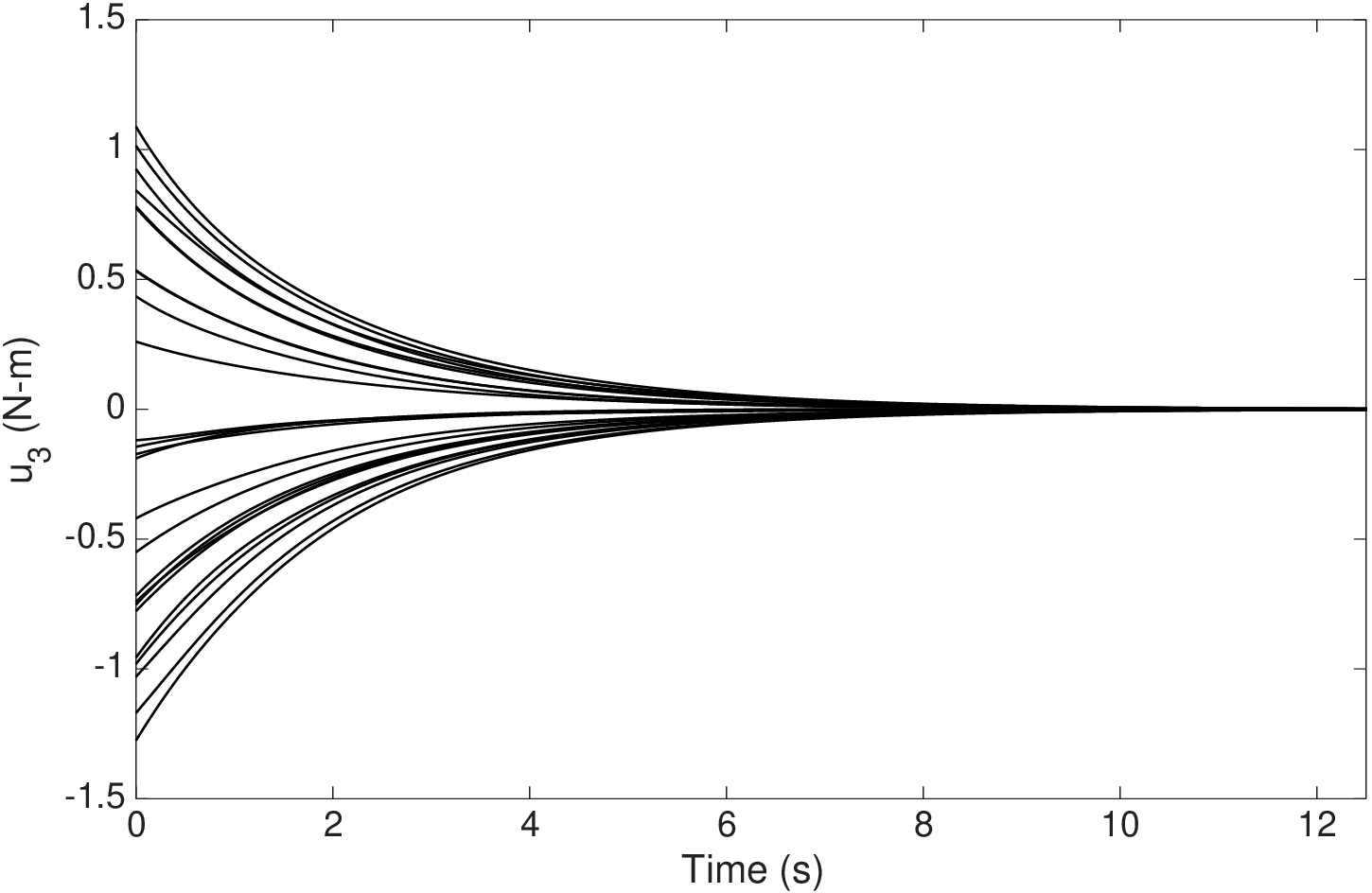}
  \end{minipage} 
  \caption{Monte Carlo Simulations of Learned Optimal Controller for Spacecraft Detumbling}
  \label{fig:detumb_state_and_control}
\end{figure*}

\begin{figure}
    \centering
    \includegraphics[width=0.35\textwidth]{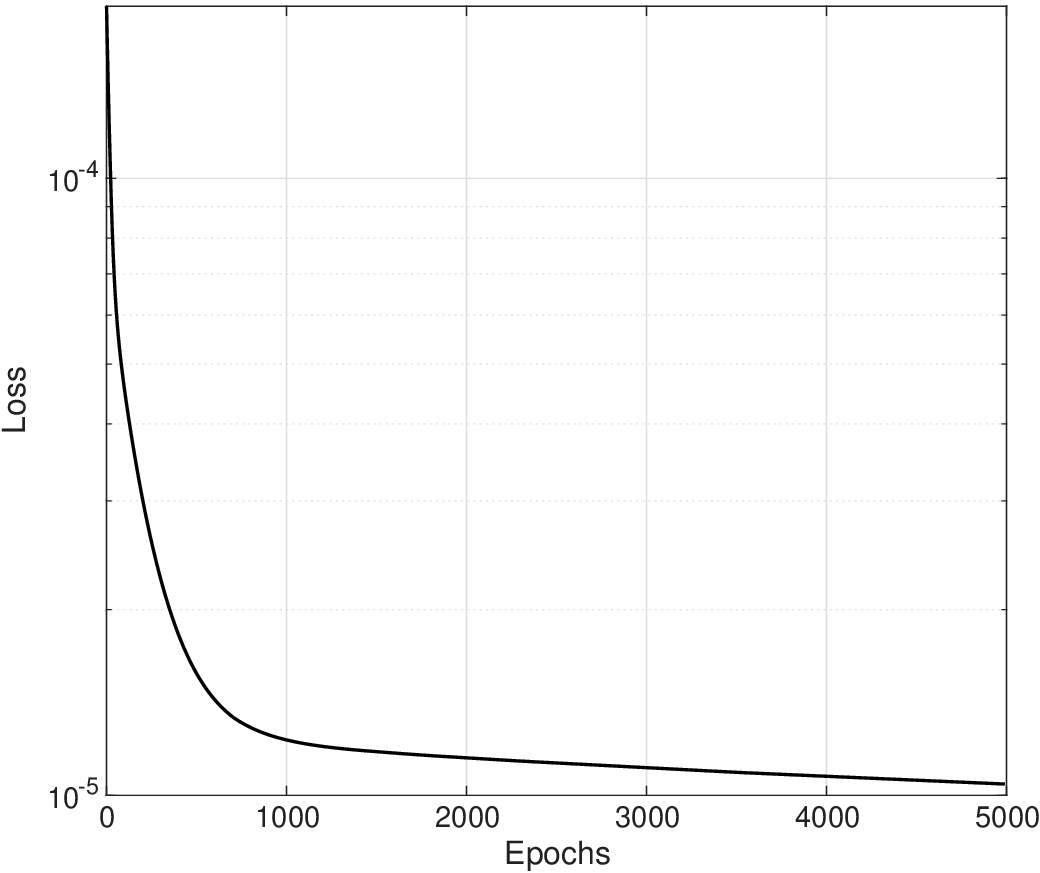}
    \caption{Loss Reduction during Training of Optimal Policy for Detumbling Problem}
    \label{fig:p3_loss}
\end{figure}

Consider the following infinite horizon satellite detumbling OCP:
\begin{equation}
    \min_{\mathbf{u}\in\mathcal{U}} J(t,\mathbf{x},\mathbf{u}) = \int_0^\infty (\mathbf{x}^\top \mathbf{Q}\mathbf{x} + \mathbf{u}^\top \mathbf{R}\mathbf{u}) dt
\end{equation}

subject to
\begin{align}
    \dot{\mathbf{x}} &= \mathbf{I}^{-1}(\mathbf{u}-\mathbf{x} \times \mathbf{I}\mathbf{x})
\end{align}

where, $\mathbf{x} \in \mathbb{R}^3$ and represents the body rate of the satellite, I is the inertia matrix of the satellite, and u is the control torque from spacecraft propulsion. Since the state space for the de-tumbling problem is higher than two dimensions, visualizing the optimal value function directly is challenging. Instead, Fig \ref{fig:detumb_state_and_control} presents results from Monte Carlo simulations, illustrating the global convergence of the learned controller within the compact state space. Fig \ref{fig:p3_loss} shows the evolution of the loss function over training epochs. Additionally, Table \ref{tab:detumbling-loss-comparison} compares model complexity with the final training loss. All results in Table \ref{tab:detumbling-loss-comparison} are obtained by training the X-TFC networks on 25000 datapoints, and 5000 training epochs.

\begin{table}[hbp]
    \caption{Detumbling Loss Comparison for Varying Hidden Layer Size}
    \centering
    \begin{tabular}{ccc}
        \hline
        \textbf{Model Name} & \textbf{Hidden Neurons} & \textbf{Final Loss} \\
        \hline
        X-TFC XS & 400 & 3.216e-05 \\
        X-TFC Small & 1000 & 5.947e-05 \\
        X-TFC Medium & 2000 & 1.043e-05 \\
        X-TFC Large & 3000 & 9.617e-05 \\
        \hline
    \end{tabular}
    \label{tab:detumbling-loss-comparison}
\end{table}


\section{Conclusions and Future Works}
This work presents a framework for synthesizing optimal feedback controllers for infinite-horizon problems by solving the HJB-PDE using the Extreme Theory of Functional Connections (X-TFC). By combining the analytical structure of TFC with the efficiency of ELM, the proposed method enables fast and accurate learning of value functions while satisfying boundary conditions exactly. We observed that the current initialization strategy based on the LQR solution does not scale well to higher-dimensional problems. In fact, for the spacecraft detumbling case, random initialization yielded better performance. In future work, we plan to explore improved initialization techniques by pre-training the value network using supervised learning on data generated from trajectory optimization. Additionally, we aim to extend the proposed X-TFC framework to solve HJB PDEs arising from finite-horizon optimal control problems.  
\label{conc}







\bibliography{references.bib}

@article{rao2009survey,
  title={A survey of numerical methods for optimal control},
  author={Rao, Anil V},
  journal={Advances in the astronautical Sciences},
  volume={135},
  number={1},
  pages={497--528},
  year={2009},
  publisher={Univelt, Inc.}
}

@INPROCEEDINGS{10883705,
  author={Joshi, Vishnu and Kumar, Suraj and V, Nithin and Kolathaya, Shishir},
  booktitle={2024 Tenth Indian Control Conference (ICC)}, 
  title={Discrete Time Model Predictive Control for Humanoid Walking with Step Adjustment}, 
  year={2024},
  volume={},
  number={},
  pages={102-107},
  doi={10.1109/ICC64753.2024.10883705}}

@article{garcia1989model,
  title={Model predictive control: Theory and practice—A survey},
  author={Garcia, Carlos E and Prett, David M and Morari, Manfred},
  journal={Automatica},
  volume={25},
  number={3},
  pages={335--348},
  year={1989},
  publisher={Elsevier}
}

@book{rawlings2017model,
  title={Model predictive control: theory, computation, and design},
  author={Rawlings, James Blake and Mayne, David Q and Diehl, Moritz and others},
  volume={2},
  year={2017},
  publisher={Nob Hill Publishing Madison, WI}
}

@book{bertsekas2012dynamic,
  title={Dynamic programming and optimal control: Volume I},
  author={Bertsekas, Dimitri},
  volume={4},
  year={2012},
  publisher={Athena scientific}
}

@article{osti_1595805,
  author       = {Raissi, Maziar and Perdikaris, Paris and Karniadakis, George Em},
  title        = {Physics-informed neural networks: A deep learning framework for solving forward and inverse problems involving nonlinear partial differential equations},
  doi          = {10.1016/j.jcp.2018.10.045},
  journal      = {Journal of Computational Physics},
  issn         = {ISSN 0021-9991},
  number       = {C},
  volume       = {378},
  place        = {United States},
  publisher    = {Elsevier},
  year         = {2018},
  month        = {11}}

@article{schiassi2020,
author = {Schiassi, Enrico and D’Ambrosio, Andrea and Drozd, Kristofer and Curti, Fabio and Furfaro, Roberto},
title = {Physics-Informed Neural Networks for Optimal Planar Orbit Transfers},
journal = {Journal of Spacecraft and Rockets},
volume = {59},
number = {3},
pages = {834-849},
year = {2022},
doi = {10.2514/1.A35138}
}

@article{HUANG2006489,
title = {Extreme learning machine: Theory and applications},
journal = {Neurocomputing},
volume = {70},
number = {1},
pages = {489-501},
year = {2006},
note = {Neural Networks},
issn = {0925-2312},
doi = {https://doi.org/10.1016/j.neucom.2005.12.126},
author = {Guang-Bin Huang and Qin-Yu Zhu and Chee-Kheong Siew}
}

@article{SIRIGNANO20181339,
title = {DGM: A deep learning algorithm for solving partial differential equations},
journal = {Journal of Computational Physics},
volume = {375},
pages = {1339-1364},
year = {2018},
issn = {0021-9991},
doi = {https://doi.org/10.1016/j.jcp.2018.08.029},
author = {Justin Sirignano and Konstantinos Spiliopoulos}
}

@Article{make2010004,
AUTHOR = {Leake, Carl and Mortari, Daniele},
TITLE = {Deep Theory of Functional Connections: A New Method for Estimating the Solutions of Partial Differential Equations},
JOURNAL = {Machine Learning and Knowledge Extraction},
VOLUME = {2},
YEAR = {2020},
NUMBER = {1},
PAGES = {37--55},
ISSN = {2504-4990},
DOI = {10.3390/make2010004}
}

@article{schiassi2021extreme,
  title={Extreme theory of functional connections: A fast physics-informed neural network method for solving ordinary and partial differential equations},
  author={Schiassi, Enrico and Furfaro, Roberto and Leake, Carl and De Florio, Mario and Johnston, Hunter and Mortari, Daniele},
  journal={Neurocomputing},
  volume={457},
  pages={334--356},
  year={2021},
  publisher={Elsevier}
}

@misc{schiassi2020extremetheoryfunctionalconnections,
      title={Extreme Theory of Functional Connections: A Physics-Informed Neural Network Method for Solving Parametric Differential Equations}, 
      author={Enrico Schiassi and Carl Leake and Mario De Florio and Hunter Johnston and Roberto Furfaro and Daniele Mortari},
      year={2020},
      eprint={2005.10632},
       
}

@book{rockafellar1997convex,
  title={Convex analysis},
  author={Rockafellar, R Tyrrell},
  volume={28},
  year={1997},
  publisher={Princeton university press}
}

@inproceedings{lutter2020hjb,
  title={HJB optimal feedback control with deep differential value functions and action constraints},
  author={Lutter, Michael and Belousov, Boris and Listmann, Kim and Clever, Debora and Peters, Jan},
  booktitle={Conference on Robot Learning},
  pages={640--650},
  year={2020},
  organization={PMLR}
}

@inproceedings{dung_6_2023_1803,
author = {Duong V. Dung and Nguyen D. Song and Pramudita S. Palar and Lavi R. Zuhal},
title = {On The Choice of Activation Functions in Physics-Informed Neural Network for Solving Incompressible Fluid Flows},
booktitle = {AIAA SCITECH 2023 Forum},
chapter = {},
pages = {},
doi = {10.2514/6.2023-1803},
year = {2023}
}

\end{document}